\documentclass[12pt,preprint]{aastex}

\slugcomment{}
\shorttitle{Eight new lenses from SQLS}
\shortauthors{}

\begin{document}

\title{Eight New Quasar Lenses from the Sloan Digital Sky Survey Quasar
Lens Search}

\author{
Issha Kayo\altaffilmark{1,2,3},
Naohisa Inada\altaffilmark{4},
Masamune Oguri\altaffilmark{5,6},
Tomoki Morokuma\altaffilmark{3,5},
Patrick B. Hall\altaffilmark{7},
Christopher S. Kochanek\altaffilmark{8},
Donald P. Schneider\altaffilmark{9}
}

\altaffiltext{1}{Institute for the Physics and Mathematics of the
Universe, University of Tokyo, 5-1-5 Kashiwanoha, Chiba 277-8582, Japan}
\altaffiltext{2}{Institute of Cosmology and Gravitation, University of Portsmouth, Burnaby Road, Portsmouth, PO1 3FX, UK}
\altaffiltext{3}{Research Fellow of the Japan Society for the Promotion of Science}
\altaffiltext{4}{Cosmic Radiation Laboratory, RIKEN, 2-1 Hirosawa, Wako, Saitama 351-0198, Japan.} 
\altaffiltext{5}{National Astronomical Observatory, Osawa, Mitaka, Tokyo 181-8588, Japan}
\altaffiltext{6}{Kavli Institute for Particle Astrophysics and 
  Cosmology, Stanford University, 2575 Sand Hill Road, Menlo Park, 
  CA 94025, USA}
\altaffiltext{7}{Department of Physics and Astronomy, York University, 4700 Keele Street, Toronto, Ontario M3J 1P3, Canada}
\altaffiltext{8}{Department of Astronomy, The Ohio State University, 4055 McPherson Lab, 140 West 18th Avenue,
Columbus, OH 43210, USA}
\altaffiltext{9}{Department of Astronomy and Astrophysics, Pennsylvania
State University, 525 Davey Laboratory, University Park, PA
16802, USA}

\begin{abstract}
  We report the discovery and confirmation of eight new two-image
  lensed quasars by the Sloan Digital Sky Survey (SDSS) Quasar Lens
  Search. The lenses are SDSS~J0904+1512 (image separation
  $\theta=1\farcs13$, source redshift $z_s=1.826$), SDSS~J1054+2733
  ($\theta=1\farcs27$, $z_s=1.452$), SDSS~J1055+4628
  ($\theta=1\farcs15$, $z_s=1.249$), SDSS~J1131+1915
  ($\theta=1\farcs46$, $z_s=2.915$), SDSS~J1304+2001
  ($\theta=1\farcs87$, $z_s=2.175$), SDSS~J1349+1227
  ($\theta=3\farcs00$, $z_s=1.722$), SDSS~J1455+1447
  ($\theta=1\farcs73$, $z_s=1.424$), and SDSS~J1620+1203
  ($\theta=2\farcs77$, $z_s=1.158$).  Three of them, SDSS~J1055+4628,
  SDSS~J1455+1447, and SDSS~J1620+1203, satisfy the criteria for
  constructing our statistical sample for studying the cosmological
  model.  Based on galactic absorption lines of the lens galaxies, we
  also derive lens redshifts of $z_l=0.398$ and $z_l=0.513$ for
  SDSS~J1620+1203 and the previously discovered lens SDSS~J0746+4403,
  respectively.
\end{abstract}


%%%%%%%%%%%%%%%%%%%%%%%%%%%%%%%%%%%%%
\section{Introduction}\label{sec:intro}
%%%%%%%%%%%%%%%%%%%%%%%%%%%%%%%%%%%%%
Gravitationally lensed quasars play important roles not only in
investigating the physical properties of lens galaxies but also in
extracting cosmological information and constraining the structure of
quasar accretion disks \citep[e.g., see][for
reviews]{schneider06,kochanek06}.  To use lensed quasars for these
studies, it is important to construct large samples, especially for
cosmological purposes.  The Cosmic Lens All-Sky Survey
\citep[CLASS;][]{myers03,browne03} is one of the largest, with 22
lenses drawn from a sample of $\sim16,000$ radio sources, and has
provided useful constraints on the evolution of galaxies and
cosmology.  We are performing a systematic lensed quasar survey, the
Sloan Digital Sky Survey Quasar Lens Search
\citep[SQLS;][]{oguri+06,inada08}, to construct a larger sample of
lensed quasars in the optical.  Thus far, we have discovered 28 lensed
quasars (26 are galaxy-scale lenses and two are cluster-scale lenses)
and rediscovered 11 known lenses \citep[e.g.,][and references
therein]{oguri+08,inada09a}.  
Our current sample is now large enough for the statistical errors on
cosmological parameters to be comparable to the present level of the
systematic uncertainties \citep{oguri08III}.
Larger samples of lensed quasars also allow
an increasing level of ``self-calibration'' to constrain many of these
uncertainties, particularly the velocity dispersion function of the
lens galaxies and its evolution as one of the largest contributors
to these systematic uncertainties \citep[e.g.,][]{choi07,
  matsumoto08, chae08}.

In this paper, we report the discovery of new eight galaxy-scale lenses
done by early in 2009, bringing our total sample to 47\footnote{Visit
  http://www-utap.phys.s.u-tokyo.ac.jp/\~{}sdss/sqls/ for a list of
  lensed quasars in the SQLS.}.  We briefly describe the selection of
lens candidates from the Sloan Digital Sky Survey
\citep[SDSS;][]{york00} in the following section, and then present the
imaging and spectroscopic observations needed to confirm these
candidates in Section~\ref{sec:follow-up}. Simple mass models of the eight
lenses are made in Section~\ref{sec:model} to see whether the lensing
interpretation is reasonable.  Throughout the paper, we assume a
standard cosmological model with matter density $\Omega_M=0.26$,
cosmological constant $\Omega_\Lambda=0.74$, and Hubble constant
$H_0=72 {\rm km}\ {\rm s}^{-1} {\rm Mpc}^{-1}$.

%%%%%%%%%%%%%%%%%%%%%%%%%%%%%%%%%%%%%
\section{SDSS Data and Selection of Candidates}\label{sec:sdss}
%%%%%%%%%%%%%%%%%%%%%%%%%%%%%%%%%%%%%

The SDSS-I and SDSS-II Sloan Legacy Surveys are photometric and
spectroscopic surveys
\citep{fukugita96,gunn98,lupton99,stoughton02,blanton03,tucker06}
covering a quarter of the all sky, using a dedicated telescope
\citep{gunn06} at the Apache Point Observatory in New Mexico, USA.
The imaging data were processed by the photometric pipeline
\citep{lupton01} and carefully calibrated
\citep{hogg01,smith02,pier03,ivezic04}. The spectroscopic quasar
targets were selected from the imaging data according to the algorithm
described by \cite{richards02} and cataloged in
\cite{schneider07} and D. P. Schneider et al. (2010, in preparation).  All the SDSS data are publicly
available in the final Data Release 7 \citep{abazajian09}.

\begin{figure}
\includegraphics[scale=0.5]{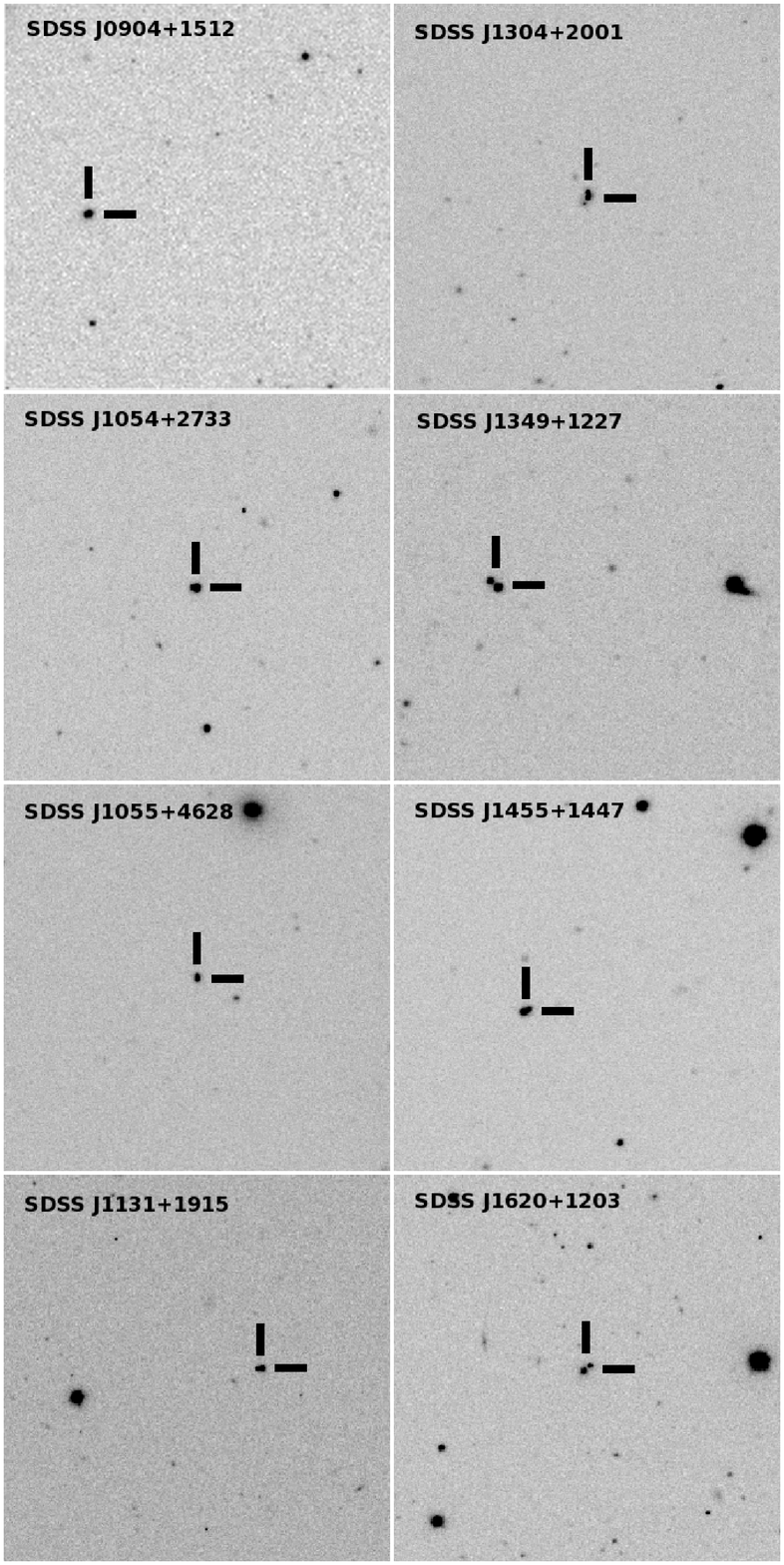}
\caption{
SDSS $i$-band images of the eight lens systems. The size of each image
 is $2'\times 2'$, and north is up and east is left.
\label{fig:sdssi}}
\end{figure}

From the spectroscopically confirmed quasar catalogs we select
candidates for lensed quasars using two different methods based on
morphologies (morphological selection) and color (color selection).
The details of the two selection methods are found in \citet{inada08}
and \citet{oguri+06}. Morphological selection is used to find lenses
with small image separations, $\lesssim 2\farcs5$, which are not
deblended by the SDSS pipeline, and color selection is for lenses with
larger separations.  Of the eight systems reported in this paper, five
systems (SDSS~J0904+1512, SDSS~J1054+2733, SDSS~J1055+4628,
SDSS~J1131+1915, and SDSS~J1455+1447) were morphologically selected,
and three (SDSS~J1304+2001, SDSS~J1349+1227, and SDSS~J1620+1203) were
color selected.  The finding charts for the systems are shown in
Figure~\ref{fig:sdssi}, and the SDSS properties of the systems are listed
in Table~\ref{tbl:sdss}.  Three systems, SDSS~J1055+4628,
SDSS~J1455+1447, and SDSS~J1620+1203, satisfy the criteria for our
statistical sample used for cosmological studies \citep[source
redshift $0.6<z_s<2.2$, image magnitude $i<19.1$, image separation
$1''<\theta<20''$, flux ratio between images $\Delta I$ or $\Delta
i<1.25$; see][for the complete description]{inada08,oguri08III,inada09a}.

%%%%%%%%%%%%%%%%%%%%%%%%%%%%%%%%%%%%%
\section{Additional Imaging and Spectroscopy}\label{sec:follow-up}
%%%%%%%%%%%%%%%%%%%%%%%%%%%%%%%%%%%%%
The imaging observations for these lenses were performed
using four instruments on the University of Hawaii 2.2 m telescope
(UH88): the Tektronix 2048x2048 CCD camera (Tek2k; the pixel size is
$0\farcs2195$ pixel$^{-1}$), the UH8k wide-field imager (UH8k;
$0\farcs235$ pixel$^{-1}$), the Orthogonal Parallel Transfer Imaging
Camera (OPTIC; $0\farcs1374$ pixel$^{-1}$), and the Quick Infrared
Camera (QUIRC; $0\farcs189$ pixel$^{-1}$).  The spectroscopic data and
some images were taken with the Faint Object Camera and Spectrograph
\citep[FOCAS;][]{kashikawa02} on the Subaru telescope. The 
FOCAS data were binned $2\times2$ on the detector leading to a
spatial resolution of $0\farcs208$ pixel$^{-1}$.  The spectral
resolution is $R\sim 400-500$. Tables~\ref{tbl:followup1} and
\ref{tbl:followup2} summarize the
observations.

\begin{figure}
\plotone{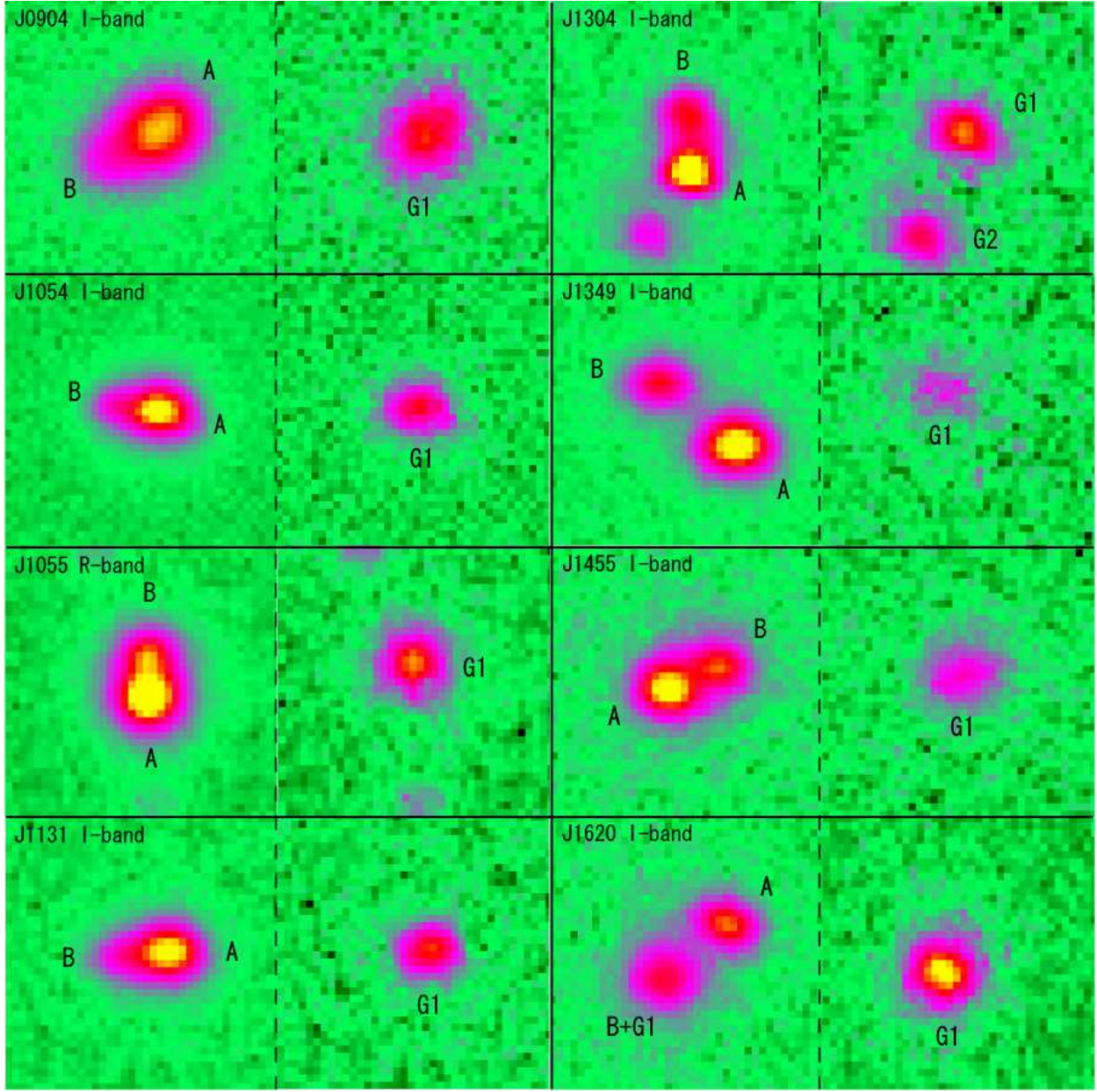}
\caption{
Images taken by the UH88 and Subaru telescopes (left panels of each)
and the residuals after fitting and subtracting two PSFs using GALFIT (right panels of each).
The residual objects labeled by ``G1'' or ``G2'' are the lens galaxies. 
These are Tek2k $I$-band images except for SDSS~J0904+1512 ($I$ band, OPTIC) and SDSS~J1055+4628 ($R$ band, FOCAS). The size of the images is approximately $8''\times8''$, and north is up and east is left.
\label{fig:allimg}}
\end{figure}

We analyzed the imaging data using GALFIT \citep{peng02}.  First we
fit each system with two stellar components using stars near the
systems as point-spread function (PSF) templates. There remained
significant extended residuals between the point sources after
subtracting the best-fit model for all systems. We then added a galaxy
modeled by a S\'ersic profile to the fit and found virtually no
residuals.  In the left panels of Figure~\ref{fig:allimg}, we show the
original $I$-band images ($R$-band image for SDSS~J1055+4628). Most of
the signal in the images arises from two point-like objects, labeled
``A'' and ``B''.  In the right panels we show the residuals, labeled
``G1'' or ``G2'', after subtracting the two best-fit PSFs.  The
astrometry and photometry of the components are listed in
Table~\ref{tbl:astro}, and the best-fit S\'ersic parameters are shown
in Table~\ref{tbl:galmodel}.  The differences in the relative
positions of these components between bands are below $0\farcs05$ for
the stellar components and $0\farcs2$ for the extended components.

\begin{figure}
\plotone{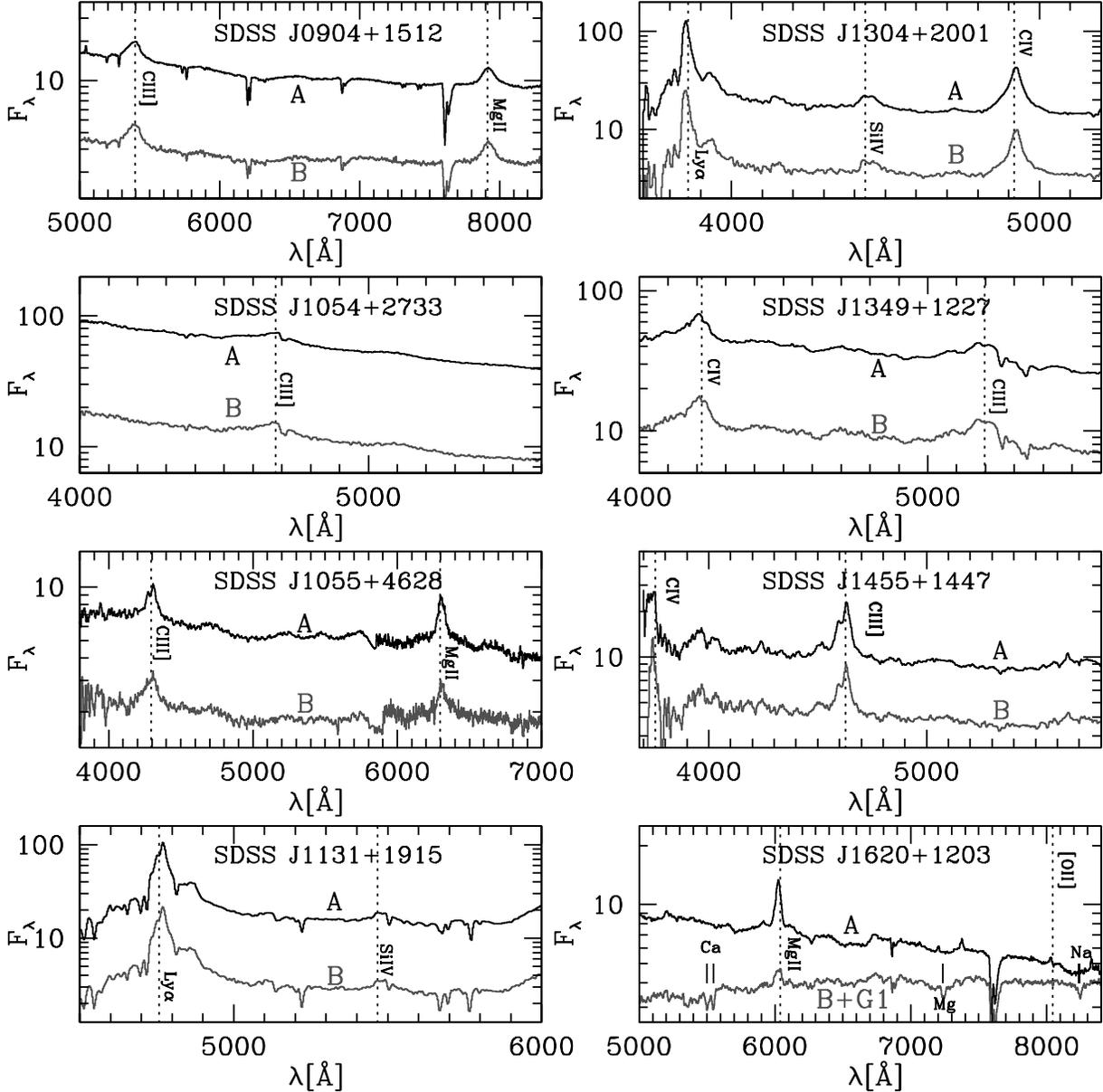}
\caption{
FOCAS spectra of the eight lens systems. The flux density units are $10^{-17}{\rm erg}\ {\rm cm}^{-2}{\rm s}^{-1}{\rm \AA}^{-1}$.
For SDSS~J1055+4628, two exposures were merged at $\sim$5850\AA.
Absorption lines from the lens galaxy are observed in the spectrum of
the fainter image of SDSS~J1620+1203. Absorption lines by some 
intervening systems are also observed in 
SDSS~J0904+1512 (\ion{Mg}{2} and \ion{Fe}{2} at $z=1.2168\pm0.0002$), 
SDSS~J1054+2733 (\ion{Mg}{2} and \ion{Fe}{2} at $z=0.6794\pm0.0002$), 
SDSS~J1131+1915 (\ion{Fe}{2} at $z=1.1890\pm0.0005$ 
and a DLA at $z=2.6025\pm0.0002$ with \ion{Si}{2}, \ion{O}{1}, and \ion{C}{2}), 
and SDSS~J1349+1227 (\ion{Fe}{2} at $z=1.2347\pm0.0002$ and $z=1.2385\pm0.0002$).
The features at $\sim$6900\AA\  and $\sim$7600\AA\  are telluric.
\label{fig:allspec}}
\end{figure}

The one-dimensional spectra of the stellar components were extracted
using standard IRAF\footnote{IRAF is distributed by the National
  Optical Astronomy Observatories, which are operated by the
  Association of Universities for Research in Astronomy, Inc., under
  cooperative agreement with the National Science Foundation.} tasks.
The spectra shown in Figure \ref{fig:allspec} and the results of the
imaging observations unambiguously confirm that the eight systems are
gravitationally lensed systems.  The shapes of the various emission
lines and continua are almost perfectly identical between the two
quasars of each system. The spectrum of the fainter quasar of
SDSS~J1620+1203 is contaminated with the relatively bright lens
galaxy, and we clearly detect some galactic absorption lines of the
lens galaxy.  We conclude with comments on
individual lens system.

\paragraph{SDSS~J0904+1512.}
The $I$-band image shows that the lens galaxy is located near the
brighter lens image, like HE1104-1805 \citep{wisotzki93} or
SDSS~J1226-0006 \citep{inada08}, and a fit to the galaxy profile
yields the S\'ersic index of 4.1.  When analyzing $V$- and $R$-band
images, we fixed the galaxy profile to the $I$-band profile because
the galaxy is too faint to fit the galaxy profile correctly.  The
image separation is $\theta=1\farcs128\pm0\farcs007$, and the source
redshift is $z_s=1.826\pm0.002$.  This lens system is not included in
the SQLS statistical sample because of the large flux ratio between
the quasar images \citep[$\Delta I > 1.25$;][]{inada08}.

\paragraph{SDSS~J1054+2733.}
The best-fit S\'ersic index to the galaxy is 3.7 in the $I$-band, and the
lens galaxy is likely to be early-type. For the $R$-band analysis, we
fixed the galaxy profile to that from the $I$-band fit. Although there remain
some residuals in the 2PSFs model of the $V$-band image, we could not
fit the residuals well because of a lack of bright enough PSF
templates to fit the weak residuals.  The image separation is
$\theta=1\farcs269\pm0\farcs004$ and the source redshift is
$z_s=1.452\pm0.002$.  This system is not included in the SQLS
statistical sample because of the large flux ratio between the images.

\paragraph{SDSS~J1055+4628.}
The FOCAS $R$-band image was used for astrometric measurements because
it is the deepest.  In addition to the two quasars (A and B) and the
lens galaxy (G1 with a best-fit S\'ersic index of 5.4), we also found
two extended objects located $3\farcs5$ to the north and $4''$ to the
south of the system. The northern object has $R-I\sim 1.0$, which is
similar to the lens galaxy G1 ($R-I=0.87$), but the southern one is
much bluer than the two galaxies.  In the $V$-band image we could not
find any galaxy components.  The image separation is
$\theta=1\farcs146\pm0\farcs004$ and the source redshift is
$z_s=1.249\pm0.001$.  This system satisfies our statistical sample
criteria and is included in SQLS statistical sample from DR5
(N. Inada et al. 2010, in preparation).

\paragraph{SDSS~J1131+1915.}
The lens galaxy was successfully fit in the $I$ band image with the
S\'ersic profile (the best-fit index is 2.4), but the $V$-band
residuals after subtracting two PSFs are faint and we could not
estimate the parameters of the galaxy. We did not take $R$-band
images.  The image separation is $\theta=1\farcs462\pm0\farcs007$ and
the source redshift is $z_s=2.915\pm0.001$.  This system is not
included in the SQLS statistical sample because the redshift is
outside of our selection criteria ($z_s<2.2$).

\paragraph{SDSS~J1304+2001.}
We fit the images with two PSFs (A and B) and two S\'ersic profiles
(G1 and G2), and the best-fit S\'ersic indices are 2.9 and 6.0,
respectively. Galaxy G1 lies between the two quasars and galaxy G2 is
located about $3\farcs5$ to the south of G1.  G1 and G2 have similar,
red colors ($R-I=0.7\sim 1.0$ and $V-R=0.8\sim0.9$), suggesting that
they are probably associated with each other.  The image separation is
$\theta=1\farcs865\pm0\farcs004$ and the source redshift is
$z_s=2.175\pm0.002$.  This system is not included in the SQLS
statistical sample because of the large flux ratio between the images.

\paragraph{SDSS~J1349+1227.}
Although this system had been reported as a binary quasar in
\citet{hennawi06}, our new observations suggest that it is a
gravitational lens rather than a binary quasar. We detected a lens
galaxy very close to the fainter quasar component with a best-fit
S\'ersic index of 1.7.  Because the galaxy is faint in the $R$- and $V$-bands, we fixed the galaxy parameters to the best-fit values from the
$I$-band. The image separation is $\theta=3\farcs002\pm0\farcs004$ and
the source redshift is $z_s=1.722\pm0.002$.  This system is not
included in the SQLS statistical sample because the flux ratio is
slightly outside of our selection criteria (N. Inada et al. 2010, in preparation).

\paragraph{SDSS~J1455+1447.}
We fit the $I$-band image with two PSFs and a galaxy,
and the best-fit S\'ersic index was 3.5. As in the case of
SDSS~J1349+1227, we fixed the galaxy profile in $R$ and $V$ to the
best-fit values from the $I$-band.  The image separation is
$\theta=1\farcs727\pm0\farcs004$ and the source redshift is
$z_s=1.424\pm0.001$.  This system will be included in the SQLS DR7
statistical sample (N. Inada et al. 2010, in preparation).

\paragraph{SDSS~J1620+1203.}
The lens galaxy is bright and close to the faint quasar, and the
spectral features of the galaxy are seen in the spectrum of the faint
quasar. In fact, the fainter quasar image was classified as a galaxy in
the SDSS due to the bright lens galaxy. \ion{Ca}{2} H\&K , Mg, and Na
galactic absorption lines are observed, giving a lens galaxy redshift
of $z_l=0.398\pm0.001$.  The image separation is
$\theta=2\farcs765\pm0\farcs011$ and the source redshift is
$z_s=1.158\pm0.002$.  The best-fit S\'ersic index of $6.62\pm2.53$ is
larger than is expected \citep{blanton05}, but we do not take it
seriously because of the large error.  This system will be included in
the SQLS DR7 statistical sample (N. Inada et al. 2010, in preparation).

%%%%%%%%%%%%%%%%%%%%%%%%%%%%%%%%%%%%%
\section{Mass Modeling}\label{sec:model}
%%%%%%%%%%%%%%%%%%%%%%%%%%%%%%%%%%%%%
We modeled all eight systems to see whether the lensing hypothesis is
reasonable from the theoretical point of view. Because of the small
number of observational constraints for two image lenses, we limited
the model to a singular isothermal ellipsoid (SIE) without any
external shear. This mass model has five parameters: the lens
position, the Einstein radius $R_E$, ellipticity $e$, and position
angle $\theta_e$ (measured east of north).  If we fit the relative
positions and the image flux ratio, these two image lenses provide
only five constraints, so our model has no degrees of freedom and we
can find a perfectly fitting model with $\chi^2\sim 0$ as long as the
model is reasonable.  To find the best-fit mass models, we used the
{\it glafic} software (M. Oguri, in preparation; version 1.0). We used
the positions and $I$-band fluxes in Table~\ref{tbl:astro}.  Only one
lens galaxy was considered even if there are nearby galaxies that
could appreciably affect the lens potentials, as in the case of
SDSS~J1055+4628 and SDSS~J1304+2001.  The resulting parameters are
summarized in Table~\ref{tbl:massmodel}.  As expected, the fitting was
done with $\chi^2\sim0$ for all lens systems.

We estimated likely redshifts of the lens galaxies from the observed
$I$-band magnitudes using the modified Faber--Jackson relation of
\cite{rusin03}. The estimated redshifts and the expected $V$- and $R$-band magnitudes at the redshifts are listed in Table \ref{tbl:fj}. For
SDSS~J1620+1203 we list the spectroscopically confirmed redshift.
Considering the scatter of the relation, $\sim 0.5$ mag, the predicted
magnitudes agree well with the observations, except for
SDSS~J0904+1512 and SDSS~J1349+1227.  The lens galaxy of SDSS
J0904+1512 is bright in the $I$-band, which makes the expected
redshift lower than suggested by the colors.  The red colors of the
lens galaxy, $R-I=1.13$ and $V-R=1.46$, suggest that it is an
early-type galaxy at $z\sim 0.5$ \citep{fukugita95}.  For
SDSS~J1349+1227, luminosity profile indicates that it might be a
late-type galaxy, which may cause the discrepancy. For comparison we
also list redshifts estimated by matching $R-I$ color of the lens
galaxies to the \cite{coleman80} early-type galaxy template. The results
are fairly consistent with the estimates from the Faber--Jackson
relation except for SDSS~J1455+1447, where the disagreement would be
solved if observed $R$-band magnitude were $\sim 0.5$ brighter.  We did
not pursue these discrepancies further because the image quality is
not adequate for a detailed investigation.
In Table~\ref{tbl:massmodel} we list the predicted time delays
using the Faber--Jackson redshift estimates from Table \ref{tbl:fj}
and the measured redshift for SDSS~J1620+1203.

%%%%%%%%%%%%%%%%%%%%%%%%%%%%%%%%%%%%%
\section{Summary}\label{sec:sum}
%%%%%%%%%%%%%%%%%%%%%%%%%%%%%%%%%%%%%
As part of the SQLS project we discovered eight new gravitationally lensed
quasars: SDSS~J0904+1512, SDSS~J1054+2733, SDSS~J1055+4628,
SDSS~J1131+1915, SDSS~J1304+2001, SDSS~J1349+1227, SDSS~J1455+1447,
and SDSS~J1620+1203.  All eight lenses are two-image quasar lenses
produced by galaxy-scale lens potentials. They were confirmed to be
lenses by imaging and spectroscopic observations using the UH88
and Subaru telescopes.  Simple mass models also suggest that the
observed image configurations and fluxes are reasonable for
lens systems.  For SDSS~J1620+1203, the redshift of the lens galaxy
was determined from the absorption lines in the spectrum of the fainter
quasar.  The system configuration is summarized in
Table~\ref{tbl:astro}.

Adding the eight systems reported in this paper, the SQLS has
discovered 36 quasar lenses and rediscovered 11 known lenses.  Among the
eight new lenses, SDSS~J1055+4628 will be included in the DR5 statistical lens
catalog (N. Inada et al. 2010, in preparation), and SDSS~J1455+1447 and SDSS~J1620+1203
will be in the final DR7 statistical lens catalog of the SQLS (N. Inada et
al. 2010, in preparation).

\acknowledgments

Use of the UH 2.2 m telescope for the observations is supported by the
National Astronomical Observatory of Japan. This work is based in part
on data collected at Subaru Telescope, which is operated by NAOJ.
I.~K. acknowledges support by Grant-in-Aid for Scientific Research on
Priority Areas No. 467 and World Premier International Research Center
Initiative (WPI Initiative), MEXT, Japan.  N.~I. acknowledges support
from the Special Postdoctoral Researcher Program of RIKEN, the RIKEN
DRI Research Grant, and MEXT KAKENHI 21740151.  This work was
supported in part by Department of Energy contract
DE-AC02-76SF00515. I.~K. and T.~M. are financially supported by the
Japan Society for the Promotion of Science Research
Fellowship. C.~S.~K. is supported by NSF grant AST-0708082.

Funding for the SDSS and SDSS-II has been provided by the Alfred
P. Sloan Foundation, the Participating Institutions, the National
Science Foundation, the U.S. Department of Energy, the National
Aeronautics and Space Administration, the Japanese Monbukagakusho, the
Max Planck Society, and the Higher Education Funding Council for
England. The SDSS Web Site is http://www.sdss.org/.

The SDSS is managed by the Astrophysical Research Consortium for the
Participating Institutions. The Participating Institutions are the
American Museum of Natural History, Astrophysical Institute Potsdam,
University of Basel, University of Cambridge, Case Western Reserve
University, University of Chicago, Drexel University, Fermilab, the
Institute for Advanced Study, the Japan Participation Group, Johns
Hopkins University, the Joint Institute for Nuclear Astrophysics, the
Kavli Institute for Particle Astrophysics and Cosmology, the Korean
Scientist Group, the Chinese Academy of Sciences (LAMOST), Los Alamos
National Laboratory, the Max-Planck-Institute for Astronomy (MPIA), the
Max-Planck-Institute for Astrophysics (MPA), New Mexico State
University, Ohio State University, University of Pittsburgh, University
of Portsmouth, Princeton University, the United States Naval
Observatory, and the University of Washington.

\appendix

%%%%%%%%%%%%%%%%%%%%%%%%%%%%%%%%%%%%%%%%%%%%%%%%%%%%%%%%%%%%%%%%%%%%%%%%%%
%%%%%%%%%%%%%%%%%%%%%%%%%%%%%%%%%%%%%%%%%%%%%%%%%%%%%%%%%%%%%%%%%%%%%%%%%%
\section{LENS REDSHIFT OF SDSS~J0746+4403}\label{sec:0746G}
%%%%%%%%%%%%%%%%%%%%%%%%%%%%%%%%%%%%%%%%%%%%%%%%%%%%%%%%%%%%%%%%%%%%%%%%%%
%%%%%%%%%%%%%%%%%%%%%%%%%%%%%%%%%%%%%%%%%%%%%%%%%%%%%%%%%%%%%%%%%%%%%%%%%%

\begin{figure}
\epsscale{.35}
\plotone{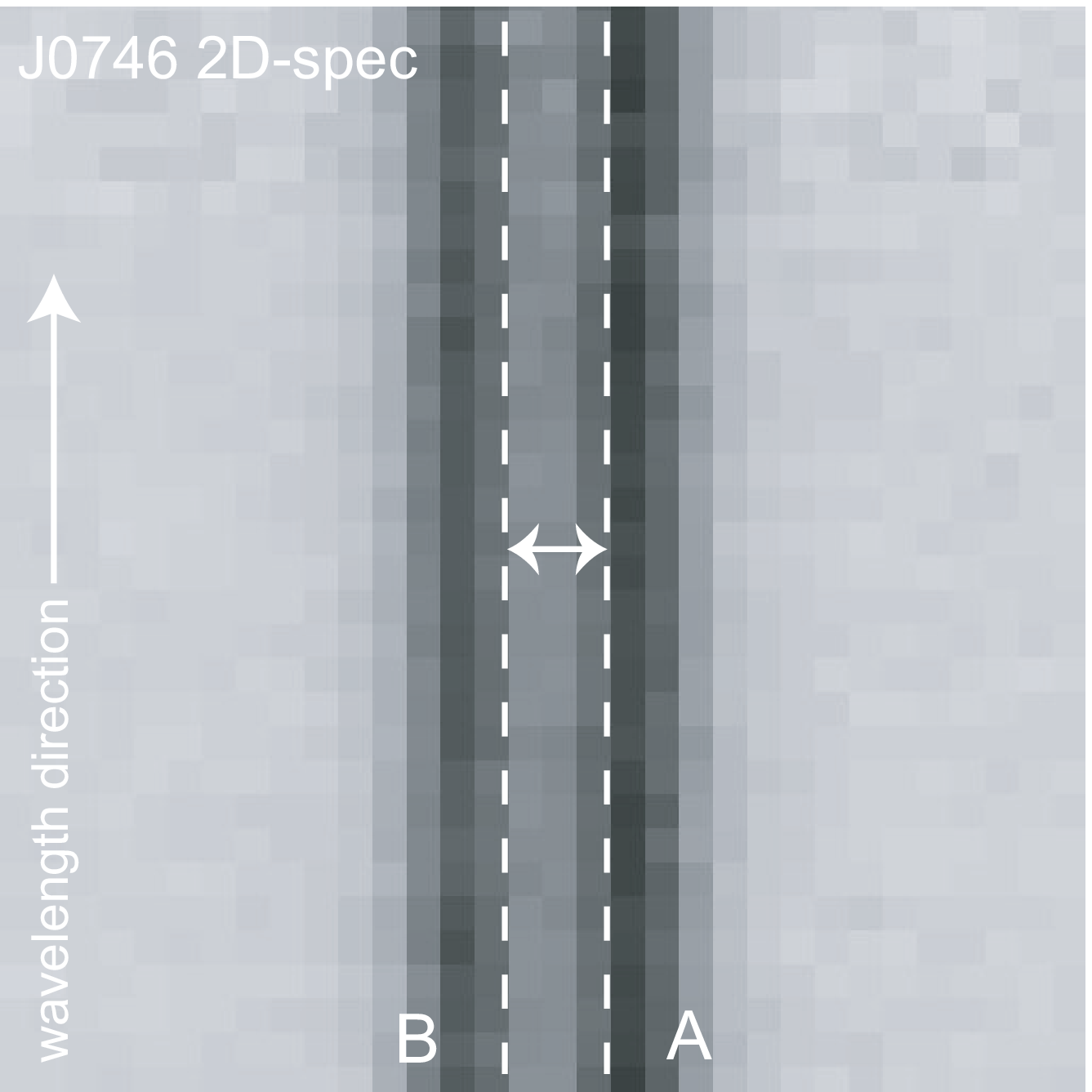}
\plotone{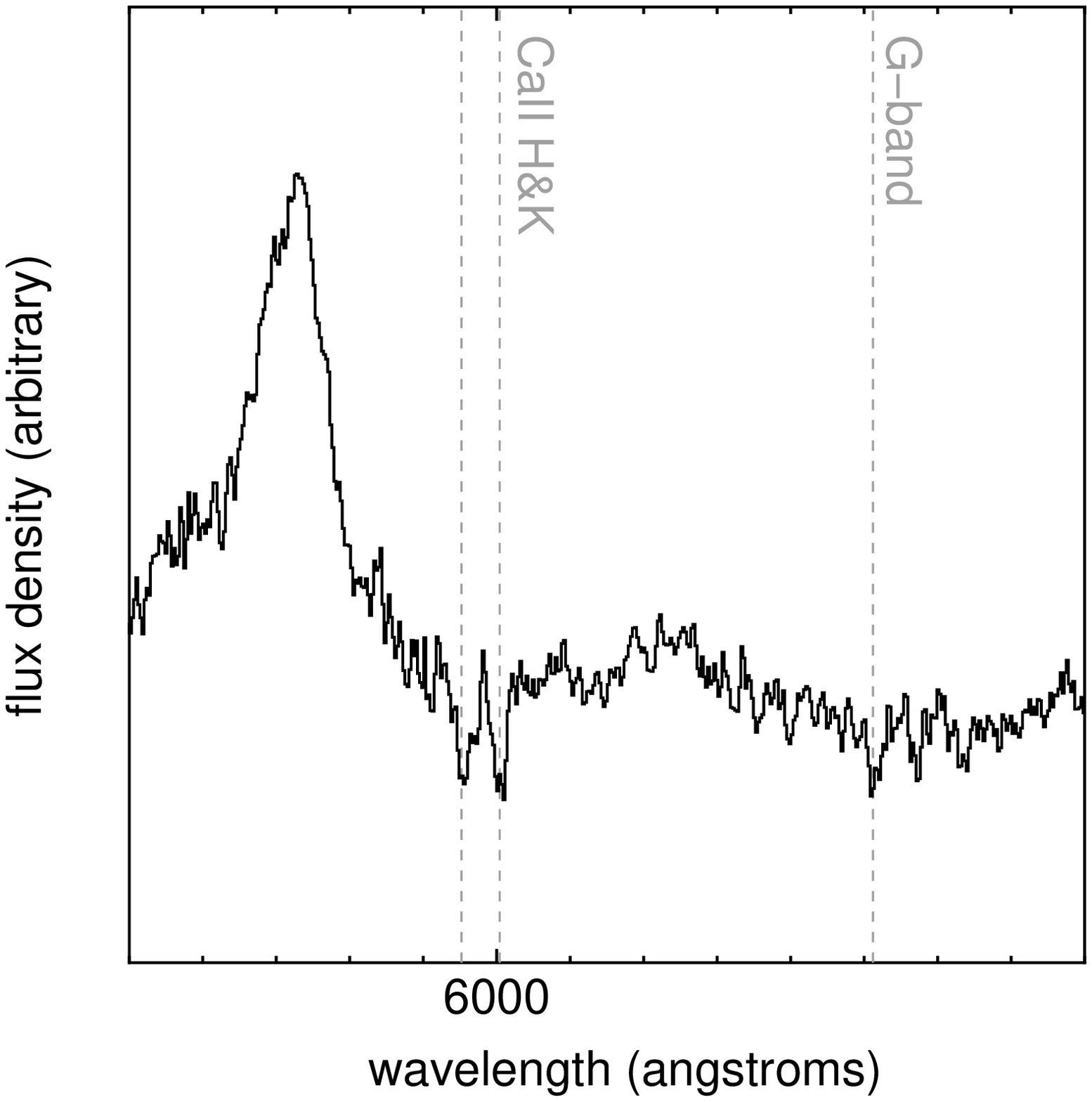}
\caption{Section of the two-dimensional FOCAS spectrum ({\em left}) of 
SDSS~J0746+4403 \citep{inada07} and 
the one-dimensional spectrum ({\em right}) extracted from the central
region shown in the left panel,
denoted by the two dashed lines. We can clearly see the
\ion{Ca}{2} H\&K and $G$-band absorption lines from the lens 
galaxy at $z_l=0.513$, in the right panel.
\label{fig:0746G}}
\end{figure}

We also measured the lens redshift of SDSS~J0746+4403 \citep{inada07}, one
of the lensed quasars included in our statistical lens sample, using the
Subaru telescope. We obtained a deep (total 2700 s exposure)
spectrum of this lens with FOCAS on 21st January 2007. We used the
300B grism, the SY47 filter and a $1\farcs0$-width slit aligned along
the A and B lensed images. The data were binned 2$\times$2 on-chip. The
seeing was less than 1\farcs0 during the exposure. The two-dimensional
spectrum of the system is shown in the left panel of
Figure~\ref{fig:0746G}.  In order to minimize the influence of the
lensed images, we extracted a one-dimensional spectrum between the peaks
of lensed images A and B (see Figure~\ref{fig:0746G}), using
standard IRAF tasks.  Although there remains some contamination from
the lensed quasar images, we can clearly see galactic absorption lines
of the lensing galaxy.  The lens redshift is measured to be
$z_l=0.513$ from the redshifted \ion{Ca}{2} H\&K and $G$-band absorption
lines at about 5950{\AA}, 6000{\AA}, and 6500{\AA}, respectively.

\clearpage

%%%%%%%%%%%%%%%%%%%%%%%%%%%%%%%%%%%%%
%%%%%%%%%%%%%%%%%%%%%%%%%%%%%%%%%%%%%

%%%%%%%%%%%%%%%%%%%%%%%%%%%%%%%%%%%%%
%%%%%%%%%%%%%%%%%%%%%%%%%%%%%%%%%%%%%

%%%%%%%%%%%%%%%%%%%%%%%%%%%%%%%%%%%%%
%%%%%%%%%%%%%%%%%%%%%%%%%%%%%%%%%%%%%

%%%%%%%%%%%%%%%%%%%%%%%%%%%%%%%%%%%%%%%%%%%%%%%%%%%%%%%%%%%%%%%%%%%%%%%
%%%%%%%%%%%%%%%%%%%%%%%%%%%%%%%%%%%%%%%%%%%%%%%%%%%%%%%%%%%%%%%%%%%%%%%

%%%%%%%%%%%%%%%%%%%%%%%%%%%%%%%%%%%%%
\begin{deluxetable}{ccccccccc}
\tabletypesize{\scriptsize}
\rotate
\tablecaption{SDSS Properties of the Lens Systems\label{tbl:sdss}}
\tablewidth{0pt}
\tablehead{ \colhead{Object} & \colhead{R.A.(J2000)} &
 \colhead{Decl.(J2000)}
 & \colhead{$u$} & \colhead{$g$} & \colhead{$r$} & \colhead{$i$} &
 \colhead{$z$}
 & \colhead{redshift}}
\startdata
SDSS~J0904+1512
 & 09:04:04.15 & +15:12:54.5
 & 17.93$\pm$0.02
 & 17.77$\pm$0.02
 & 17.80$\pm$0.01
 & 17.58$\pm$0.01
 & 17.54$\pm$0.02
 & 1.826$\pm$0.002\\ \hline
SDSS~J1054+2733
 & 10:54:40.83 & +27:33:06.4
 & 17.16$\pm$0.02
 & 16.93$\pm$0.02
 & 16.95$\pm$0.02
 & 16.84$\pm$0.02
 & 16.89$\pm$0.02
 & 1.452$\pm$0.002\\ \hline
SDSS~J1055+4628
 & 10:55:45.45 & +46:28:39.4
 & 19.38$\pm$0.04
 & 19.28$\pm$0.02
 & 18.79$\pm$0.03
 & 18.78$\pm$0.02
 & 18.79$\pm$0.04
 & 1.249$\pm$0.001\\ \hline
SDSS~J1131+1915
 & 11:31:57.72 & +19:15:27.7
 & 19.90$\pm$0.04
 & 18.23$\pm$0.02
 & 18.00$\pm$0.02
 & 18.04$\pm$0.04
 & 17.85$\pm$0.03
 & 2.915$\pm$0.001 \\ \hline
SDSS~J1304+2001
 & 13:04:43.58 & +20:01:04.2
 & 18.93$\pm$0.02
 & 18.66$\pm$0.02
 & 18.70$\pm$0.03
 & 18.49$\pm$0.03
 & 18.33$\pm$0.03
 & 2.175$\pm$0.002 \\
 & 13:04:43.60 & +20:01:05.8
 & 21.08$\pm$0.32
 & 20.45$\pm$0.35
 & 19.98$\pm$0.19
 & 19.93$\pm$0.23
 & 19.44$\pm$0.12
 & \nodata  \\ \hline
SDSS~J1349+1227
 & 13:49:29.84 & +12:27:06.8
 & 17.99$\pm$0.02
 & 17.85$\pm$0.02
 & 17.79$\pm$0.02
 & 17.51$\pm$0.03
 & 17.53$\pm$0.03
 & 1.722$\pm$0.002 \\
 & 13:49:30.00 & +12:27:08.7
 & 19.70$\pm$0.22
 & 19.32$\pm$0.17
 & 19.11$\pm$0.12
 & 18.73$\pm$0.13
 & 18.62$\pm$0.13
 & \nodata  \\ \hline
SDSS~J1455+1447
 & 14:55:01.91 & +14:47:34.8
 & 19.53$\pm$0.03
 & 19.11$\pm$0.02
 & 18.61$\pm$0.01
 & 18.30$\pm$0.02
 & 18.04$\pm$0.03
 & 1.424$\pm$0.001 \\ \hline
SDSS~J1620+1203
 & 16:20:26.14 & +12:03:42.0
 & 19.82$\pm$0.04
 & 19.70$\pm$0.03
 & 19.26$\pm$0.04
 & 19.21$\pm$0.05
 & 19.23$\pm$0.09
 & 1.158$\pm$0.002 \\
 & 16:20:26.27 & +12:03:40.3
 & 21.30$\pm$0.13
 & 20.76$\pm$0.12
 & 19.90$\pm$0.08
 & 19.52$\pm$0.06
 & 19.23$\pm$0.07
 & \nodata \\
\enddata
\tablecomments{Magnitudes are PSF magnitudes without Galactic
  extinction corrections. The two images of SDSS~J1304+2001 and
  SDSS~J1349+1227 are split by the SDSS pipeline. The fainter quasar
 in SDSS~J1620+1203 is overlapped by the bright lens galaxy and
  classified as a galaxy by the SDSS software.}
\end{deluxetable}
%%%%%%%%%%%%%%%%%%%%%%%%%%%%%%%%%%%%%

%%%%%%%%%%%%%%%%%%%%%%%%%%%%%%%%%%%%%
\begin{deluxetable}{llll}
\tabletypesize{\scriptsize}
\rotate
\tablecaption{Imaging Observations\label{tbl:followup1}}
\tablewidth{0pt}
\tablehead{ \colhead{Object} & \colhead{Imager} &
\colhead{Date of Observation (UT)} & \colhead{Exposure Time (sec)} }
\startdata
SDSS~J0904+1512
 & Tek2k($VR$), OPTIC($I$)
 & 12/Apr/07($V$), 20/Jan/09($R$), 4/May/06($I$)
 & 300($V$), 600($R$), 800($I$) \\
SDSS~J1054+2733
 & Tek2k($VRI$)
 & 12/Apr/07($VR$), 16/Nov/06($I$)
 & 300($V$), 200($R$), 600($I$) \\
SDSS~J1055+4628
 & UH8k($V$), Tek2k($BI$), FOCAS($R$)
 & 13/Nov/07($B$), 18/Nov/04($V$), 1/Feb/09($R$), 12/Nov/07($I$)
 & 400($B$), 360($V$), 120($R$), 400($I$)  \\
SDSS~J1131+1915
 & Tek2k($VIz'$)
 & 12/Apr/07($V$), 16/Apr/07($Iz'$)
 & 300($V$), 800($I$), 800($z'$)\\
SDSS~J1304+2001
 & Tek2k($VRI$)
 & 7/Mar/08($VRI$)
 & 400($V$), 400($R$), 400($I$) \\
SDSS~J1349+1227
 & Tek2k($VRI$), QUIRC($H$)
 & 16/Apr/09($VRI$), 22/Feb/05($H$)
 & 400($V$), 400($R$), 400($I$), 1080($H$) \\
SDSS~J1455+1447
 & Tek2k($VRI$)
 & 16/Apr/09($VRI$)
 & 300($V$), 300($R$), 400($I$) \\
SDSS~J1620+1203
 & Tek2k($VRI$)
 & 16/Apr/09($VRI$)
 & 400($V$), 300($R$), 300($I$) \\
\enddata
\tablecomments{Date of observation is the UT date when the exposures were started.}
\end{deluxetable}
%%%%%%%%%%%%%%%%%%%%%%%%%%%%%%%%%%%%%

%%%%%%%%%%%%%%%%%%%%%%%%%%%%%%%%%%%%%%
\begin{deluxetable}{cccccccc}
\tabletypesize{\scriptsize}
\tablecaption{FOCAS Spectroscopic Observations\label{tbl:followup2}}
\rotate
\tablewidth{0pt}
\tablehead{ \colhead{Object} & \colhead{slit width} &
  \colhead{grism} & \colhead{filter} &
  \colhead{wavelength coverage (\AA)} & \colhead{$R$} 
  & \colhead{Date of Observation (UT)} & \colhead{Exposure time (sec)}}
\startdata
SDSS~J0904+1512
 & $1''$ & 300B & SY47 & 4700-9100 & 400 & 23/Jan/07 & 1000\\
SDSS~J1054+2733
 & $1''$ & 300B & L600 & 3700-6000 & 400 & 1/Feb/09 & 480\\
SDSS~J1055+4628 (red)
 & $0.8''$ & 300R & SO58 & 5800-10000 & 500 & 7/May/08 & 300\\
SDSS~J1055+4628 (blue)
 & $1''$ & 300B & L600 & 3700-6000 & 400 & 1/Feb/09 & 720\\
SDSS~J1131+1915
 & $1''$ & 300B & L600 & 3700-6000 & 400 & 1/Feb/09 & 720\\
SDSS~J1304+2001
 & $1''$ & 300B & L600 & 3700-6000 & 400 & 1/Feb/09 & 720\\
SDSS~J1349+1227
 & $1''$ & 300B & L600 & 3700-6000 & 400 & 1/Feb/09 & 480\\
SDSS~J1455+1447
 & $1''$ & 300B & L600 & 3700-6000 & 400 & 1/Feb/09 & 480\\
SDSS~J1620+1203
 & $1''$ & 300B & SY47 & 4700-9100 & 400 & 1/Feb/09 & 720\\
\enddata
\end{deluxetable}
%%%%%%%%%%%%%%%%%%%%%%%%%%%%%%%%%%%%%%

%%%%%%%%%%%%%%%%%%%%%%%%%%%%%%%%%%%%%
\begin{deluxetable}{crrccc}
\tabletypesize{\footnotesize}
\tablecaption{Relative Astrometry and Absolute Photometry of Lens Systems\label{tbl:astro}}
\tablewidth{0pt}
\tablehead{\colhead{Component} & \colhead{$\Delta$ X (arcsec)} &
 \colhead{$\Delta$ Y (arcsec)} & \colhead{$V$} &
 \colhead{$R$} & \colhead{$I$}}
\startdata
\multicolumn{6}{l}{SDSS~J0904+1512 ($\theta=1\farcs128\pm0\farcs007$)} \\
A & 0.000$\pm$0.003  & 0.000$\pm$0.003
  & 18.18$\pm$0.01 & 17.88$\pm$0.01 & 17.71$\pm$0.04 \\
B & $-0.988\pm$0.007  & $-0.544\pm$0.004
  & 20.15$\pm$0.02 & 19.89$\pm$0.04 & 19.13$\pm$0.02 \\
G1 & $-0.217\pm$0.033  & $-0.091\pm$0.018
  & 21.19$\pm$0.12 & 19.73$\pm$0.07 & 18.60$\pm$0.07 \\
\hline
\multicolumn{6}{l}{SDSS~J1054+2733 ($\theta=1\farcs269\pm0\farcs004$)}  \\
A &  0.000$\pm$0.002     & 0.000$\pm$0.002
  & 17.21$\pm$0.01 & 17.09$\pm$0.01 & 16.94$\pm$0.01 \\
B &  $-1.260\pm$0.004    & 0.149$\pm$0.004
  & 19.22$\pm$0.01 & 18.98$\pm$0.01 & 18.82$\pm$0.02 \\
G1 &  $-0.391\pm$0.042    & $-0.048\pm$0.018
  & \nodata & 19.13$\pm$0.03 & 18.75$\pm$0.05 \\
\hline
\multicolumn{6}{l}{SDSS~J1055+4628 ($\theta=1\farcs146\pm0\farcs004$)} \\
A &  0.000$\pm$0.002       & 0.000$\pm$0.002
  & 20.02$\pm$0.01 & 19.17$\pm$0.01 & 18.98$\pm$0.01 \\
B &  $-0.024\pm$0.013      & 1.146$\pm$0.004
  & 21.07$\pm$0.01 & 20.67$\pm$0.03 & 20.17$\pm$0.04 \\
G1 &  $-0.136\pm$0.033      & 0.871$\pm$0.040
  & \nodata & 20.60$\pm$0.10 & 19.73$\pm$0.05 \\
\hline
\multicolumn{6}{l}{SDSS~J1131+1915 ($\theta=1\farcs462\pm0\farcs007$)} \\
A &  0.000$\pm$0.002      & 0.000$\pm$0.002
  & 18.51$\pm$0.01 & \nodata & 17.91$\pm$0.02 \\
B &  $-1.455\pm$0.007      & $-0.145\pm$0.004
  & 20.53$\pm$0.02 & \nodata & 19.62$\pm$0.02 \\
G1 &  $-0.378\pm$0.079       & $-0.013\pm$0.020
  & \nodata & \nodata & 19.45$\pm$0.07 \\
\hline
\multicolumn{6}{l}{SDSS~J1304+2001 ($\theta=1\farcs865\pm0\farcs004$)} \\
A &  0.000$\pm$0.002     & 0.000$\pm$0.002
  & 18.70$\pm$0.01 & $18.54\pm0.01$ & 18.05$\pm$0.01  \\
B &  -0.116$\pm$0.004    & 1.861$\pm$0.004
  & 20.23$\pm$0.01 & $20.04\pm0.01$ & 19.61$\pm$0.02 \\
G1 &  0.092$\pm$0.009    & 1.225$\pm$0.015
  & 20.68$\pm$0.16 & $19.71\pm0.01$ & 18.74$\pm$0.08 \\
G2 &  -1.245$\pm$0.009   & -1.951$\pm$0.007
  & 20.42$\pm$0.14 & $19.58\pm0.23$ & 18.96$\pm$0.17 \\
\hline
\multicolumn{6}{l}{SDSS~J1349+1227 ($\theta=3\farcs002\pm0\farcs004$)} \\
A &  0.000$\pm$0.002       & 0.000$\pm$0.002
  & 17.79$\pm$0.01 & 17.59$\pm$0.01 & 17.14$\pm$0.01 \\
B &  $-2.335\pm$0.004      & 1.886$\pm$0.004
  & 19.30$\pm$0.01 & 19.01$\pm$0.01 & 18.49$\pm$0.01 \\
G1 &  $-2.105\pm$0.048      & 1.727$\pm$0.037
  & 20.99$\pm$0.09 & 20.71$\pm$0.06 & 19.35$\pm$0.05 \\
\hline
\multicolumn{6}{l}{SDSS~J1455+1447 ($\theta=1\farcs727\pm0\farcs004$)} \\
A &  0.000$\pm$0.002       & 0.000$\pm$0.002
  & 19.50$\pm$0.02 & 18.83$\pm$0.01 & 18.25$\pm$0.01 \\
B &  1.578$\pm$0.004       & 0.702$\pm$0.004
  & 20.30$\pm$0.02 & 19.83$\pm$0.02 & 19.00$\pm$0.01 \\
G1 &  0.591$\pm$0.031       & 0.400$\pm$0.015
  & 20.29$\pm$0.08 & 19.62$\pm$0.03 & 18.51$\pm$0.07 \\
\hline
\multicolumn{6}{l}{SDSS~J1620+1203 ($\theta=2\farcs765\pm0\farcs011$)} \\
A &  0.000$\pm$0.004       & 0.000$\pm$0.002
  & 20.07$\pm$0.01 & 19.12$\pm$0.01 & 18.64$\pm$0.01 \\
B &  $-2.080\pm$0.011      & $-1.822\pm$0.011
  & 21.30$\pm$0.07 & 20.73$\pm$0.10 & 19.94$\pm$0.05 \\
G1 &  $-1.756\pm$0.013      & $-1.449\pm$0.013
  & 20.10$\pm$1.16 & 18.49$\pm$0.29 & 17.59$\pm$0.45 \\
\enddata
\tablecomments{Positions of each component were derived from the $I$-band images, except for the $R$-band image used for
  SDSS~J1055+4628. The positive directions of X and Y are west and
  north, respectively. SDSS~J1131+1915 was not observed in the $R$-band. The $V$-band fluxes of the lens galaxies of SDSS~J1054+2733,
  SDSS~J1055+4628, and SDSS~J1131+1915 are too faint to be
  measured. The errors on the positions and fluxes include only
  statistical errors reported by GALFIT. The errors on the image separation
  were calculated on the assumption that the position uncertainties are all
  uncorrelated.}
\end{deluxetable}
%%%%%%%%%%%%%%%%%%%%%%%%%%%%%%%%%%%%%

%%%%%%%%%%%%%%%%%%%%%%%%%%%%%%%%%%%%%
\begin{deluxetable}{crrrr}
\tabletypesize{\footnotesize}
\tablecaption{Luminosity Profile of Lens Galaxies}
\tablewidth{0pt}
\tablehead{\colhead{Object} & \colhead{$r_e$\tablenotemark{a}($''$)} & \colhead{$n$\tablenotemark{b}} &
  \colhead{$e$\tablenotemark{c}} & \colhead{$\theta_e$\tablenotemark{d}(deg)} }
\startdata
SDSS~J0904+1512 & $0.26\pm0.03$ & $4.11\pm1.04$ & $0.49\pm0.04$ & $-20\pm5$ \\
SDSS~J1054+2733 & $0.42\pm0.04$ & $3.67\pm0.75$ & $0.42\pm0.04$ & $-88\pm5$ \\
SDSS~J1055+4628 & $0.53\pm0.05$ & $5.42\pm1.53$ & $0.07\pm0.04$ & $+28\pm22$ \\
SDSS~J1131+1915 & $0.14\pm0.04$ & $2.38\pm1.22$ & $0.82\pm0.82$ & $-15\pm11$ \\
SDSS~J1304+2001 & $0.54\pm0.02$ & $2.90\pm0.45$ & $0.42\pm0.03$ & $+58\pm4$ \\
SDSS~J1349+1227 & $1.06\pm0.13$ & $1.69\pm0.52$ & $0.24\pm0.06$ & $+71\pm11$  \\
SDSS~J1455+1447 & $1.49\pm0.20$ & $3.50\pm0.54$ & $0.28\pm0.04$ & $-76\pm5$ \\
SDSS~J1620+1203 & $3.70\pm3.20$ & $6.62\pm2.53$ & $0.25\pm0.03$ & $-21\pm4$  \\
\enddata
\tablecomments{S\'ersic parameters measured in the $I$ band images ($R$ band for SDSS~J1055+4628) using GALFIT.}
\tablenotetext{a}{Effective radius of the S\'ersic profile.}
\tablenotetext{b}{S\'ersic index.}
\tablenotetext{c}{Ellipticity.}
\tablenotetext{d}{Major axis position angle measured east of north.}
\label{tbl:galmodel}
\end{deluxetable}
%%%%%%%%%%%%%%%%%%%%%%%%%%%%%%%%%%%%%

%%%%%%%%%%%%%%%%%%%%%%%%%%%%%%%%%%%%%
\begin{deluxetable}{crrrrr}
\tabletypesize{\footnotesize}
\tablecaption{Mass Models of Lenses}
\tablewidth{0pt}
\tablehead{\colhead{Object} & \colhead{$R_{\rm Ein}$($''$)} & \colhead{$e$} &
 \colhead{$\theta_e$(deg)} & \colhead{$\mu_{\rm tot}$\tablenotemark{a}} & \colhead{$\Delta t$ \tablenotemark{b}(day)}}
\startdata
SDSS~J0904+1512 & $0.529\pm0.007$ & $0.72\pm0.05$ & $+31\pm2$ & $6.9\pm1.0$ & $9.1\pm0.6$ \\
SDSS~J1054+2733 & $0.598\pm0.004$ & $0.52\pm0.06$ & $-14\pm1$ & $12.7\pm2.2$ & $10.1\pm1.2$ \\
SDSS~J1055+4628 & $0.600\pm0.016$ & $0.37\pm0.08$ & $-49\pm7$ & $3.4\pm0.5$ & $23.4\pm2.2$ \\
SDSS~J1131+1915 & $0.680\pm0.006$ & $0.58\pm0.09$ & $+5\pm1$  & $10.2\pm1.9$ & $22.4\pm3.3$ \\
SDSS~J1304+2001 & $0.987\pm0.008$ & $0.23\pm0.02$ & $+27\pm3$ & $7.9\pm0.8$ & $24.0\pm0.9$ \\
SDSS~J1349+1227 & $1.399\pm0.010$ & $0.57\pm0.09$ & $-42\pm4$ & $2.0\pm0.1$ & $450.1\pm8.8$ \\
SDSS~J1455+1447 & $0.853\pm0.004$ & $0.16\pm0.03$ & $+10\pm2$ & $12.7\pm2.0$ & $11.0\pm1.4$ \\
SDSS~J1620+1203 & $1.353\pm0.021$ & $0.23\pm0.06$ & $+21\pm10$ & $2.8\pm0.1$ & $179.3\pm2.0$ \\
\enddata
\tablecomments{Errors are estimated from 1000 random data for each
  system with scatter in Table~\ref{tbl:astro} for positions and 0.2
  mag for image magnitudes. We used the redshifts estimated by the
  Faber--Jackson relation in Table \ref{tbl:fj} to compute the
  predicted time delays, or the measured redshift for
  SDSS~J1620+1203. We do not include the uncertainty of the redshift
  for the error estimate.}
\tablenotetext{a}{Total magnification.}
\tablenotetext{b}{Time delay.}
\label{tbl:massmodel}
\end{deluxetable}
%%%%%%%%%%%%%%%%%%%%%%%%%%%%%%%%%%%%%

%%%%%%%%%%%%%%%%%%%%%%%%%%%%%%%%%%%%%
\begin{deluxetable}{cllrr}
\tabletypesize{\footnotesize}
\tablecaption{Predicted Redshifts and Apparent $V$ and $R$ Magnitudes of
 Lens Galaxies}
\tablewidth{0pt}
\tablehead{\colhead{Object} & \colhead{$V$} & \colhead{$R$} & \colhead{Redshift (FJ)} & \colhead{Redshift ($R-I$)}}
\startdata
SDSS~J0904+1512 & 20.1 (21.2) & 19.1 (19.7) & 0.19 & 0.54 \\
SDSS~J1054+2733 & 20.3 ($\cdots$) & 19.3 (19.1) & 0.23 & --\tablenotemark{b} \\
SDSS~J1055+4628 & 21.6 ($\cdots$) & 20.5 (20.6) & 0.39 & 0.38 \\
SDSS~J1131+1915 & 21.2 ($\cdots$) & 20.1 ($\cdots$) & 0.32 & \nodata\\
SDSS~J1304+2001 & 20.5 (20.4) & 19.3 (19.6) & 0.32  & 0.46 \\
SDSS~J1349+1227 & 21.7 (21.0) & 20.3 (20.7) & 0.63 & 0.66 \\
SDSS~J1455+1447 & 20.2 (20.3) & 19.0 (19.6) & 0.27 & 0.53 \\
SDSS~J1620+1203 & 20.2 (20.1) & 18.9 (18.5) & 0.398\tablenotemark{a} & 0.41 \\
\enddata
\tablecomments{Values in parenthesis are the observed magnitude in
  Table~\ref{tbl:astro}. The typical uncertainties of redshift (FJ) and
  redshift ($R-I$) are $\sim0.1$, considering $0.5$ mag and $0.1$ mag scatter
  for luminosity and color of galaxies, respectively.}
  \tablenotetext{a}{Redshift
  of SDSS~J1620+1203 is measured spectroscopically and $I$-band
  magnitude is estimated to be 18.3 at this redshift. FJ estimate will
  be 0.27 to recover the observed $I$-band magnitude of 17.59.}
\tablenotetext{b}{$R-I$ color is too blue to match the template.}
\label{tbl:fj}
\end{deluxetable}
%%%%%%%%%%%%%%%%%%%%%%%%%%%%%%%%%%%%%


\begin{thebibliography}{}

\bibitem[Abazajian et al.(2009)]{abazajian09}
Abazajian, K. N., et al. 2009, \apjs, 182, 543

\bibitem[Blanton et al.(2003)]{blanton03} 
Blanton, M. R., Lin, H., Lupton, R. H., Maley, F. M., Young, N., 
Zehavi, I., \& Loveday, J. 2003, \aj, 125, 2276

\bibitem[Blanton et al.(2005)]{blanton05} 
Blanton, M. R., Eisenstein, D., Hogg, D. W., Schlegel, D. J.,
 \& Brinkmann, J. 2005, \apj, 629, 143

\bibitem[Browne et al.(2003)]{browne03}
Browne, I. W. A., et al. 2003, \mnras, 341, 13

\bibitem[Chae(2008)]{chae08}
Chae, K.-H. 2008, \mnras, 402, 2031

\bibitem[Choi et al.(2007)]{choi07}
Choi, Y.-Y., Park, C., \& Vogeley, M. S. 2007, \apj, 658, 884

\bibitem[Coleman, Wu, \& Weedman(1980)]{coleman80}
Coleman, G. D., Wu, C.-C., \& Weedman, D. W. 1980, \apjs, 43, 393

\bibitem[Fukugita et al.(1996)]{fukugita96}
Fukugita, M., Ichikawa, T., Gunn, J. E., Doi, M., 
Shimasaku, K., \& Schneider, D. P. 1996, \aj, 111, 1748

\bibitem[Fukugita et al.(1995)]{fukugita95}
Fukugita, M., Shimasaku, K., \& Ichikawa, T. 1995, \pasp, 107, 945

\bibitem[Gunn et al.(1998)]{gunn98} 
Gunn, J. E., et al. 1998, \aj, 116, 3040

\bibitem[Gunn et al.(2006)]{gunn06}
Gunn, J.~E., et al. 2006, \aj, 131, 2332

\bibitem[Hennawi et al.(2006)]{hennawi06}
Hennawi, J.F., et al. 2006, \aj, 131, 1

\bibitem[Hogg et al.(2001)]{hogg01}
Hogg, D. W., Finkbeiner, D. P., Schlegel, D. J., 
\& Gunn, J. E. 2001, \aj, 122, 2129

\bibitem[Inada et al.(2007)]{inada07}
Inada, N., et al. 2007, \aj, 133, 206

\bibitem[Inada et al.(2008)]{inada08}
Inada, N., et al. 2008, \aj, 135, 496

\bibitem[Inada et al.(2009a)]{inada09a}
Inada, N., et al. 2009a, \aj, 137, 4118

\bibitem[Inada et al.(2009b)]{inada09b}
Inada, N., et al. 2009b, in preparation

\bibitem[Ivezi\'{c} et al.(2004)]{ivezic04}
Ivezi\'{c}, \v{Z}., et al. 2004, Astron. Nachr, 325, 583

\bibitem[Kashikawa et al.(2002)]{kashikawa02}
Kashikawa, N., et al. 2002, \pasj, 54, 819

\bibitem[Kochanek(2006)]{kochanek06}
Kochanek, C.S., 2006, in Gravitational Lensing: Strong
Weak and Micro, Saas-Fee Advanced Course 33, G. Meylan,
P. North, P. Jetzer, eds., (Springer: Berlin) 91

\bibitem[Lupton et al.(2001)]{lupton01}
Lupton, R., Gunn, J. E., Ivezi\'c, Z., Knapp, G. R.,
Kent, S., \& Yasuda, N. 2001, in ASP Conf. Ser. 238,
Astronomical Data Analysis Software and Systems X,
ed. F. R. Harnden, Jr., F. A. Primini, and H. E. Payne
(San Francisco: ASP), 269

\bibitem[Lupton et al.(1999)]{lupton99}
Lupton, R. H., Gunn, J. E., \& Szalay, A. S. 1999, \aj, 118, 1406

\bibitem[Matsumoto \& Futamase (2008)]{matsumoto08}
Matsumoto, A. \& Futamase, T. 2008, \mnras, 384, 843

\bibitem[Myers et al.(2003)]{myers03}
Myers, S. T., et al. 2003, \mnras, 341, 1

\bibitem[Oguri et al.(2006)]{oguri+06}
Oguri, M., et al. 2006, \aj, 132, 999

\bibitem[Oguri et al.(2008a)]{oguri08III}
Oguri, M., et al. 2008a, \aj, 135, 512

\bibitem[Oguri et al.(2008b)]{oguri+08}
Oguri, M., et al. 2008b, \mnras, 391, 1973

\bibitem[Peng et al.(2002)]{peng02}
Peng, C. Y., Ho, L. C., Impey, C. D., \& 
Rix, H.-W. 2002, \aj, 124, 266  

\bibitem[Pier et al.(2003)]{pier03} 
Pier, J. R., Munn, J. A., Hindsley, R. B., Hennessy, G. S., 
Kent, S. M., Lupton, R. H., \& Ivezi\'{c}, \'{Z}. 2003, 
\aj, 125, 1559

\bibitem[Richards et al.(2002)]{richards02} 
Richards, G. T., et al. 2002, \aj, 123, 2945

\bibitem[Rusin et al.(2003)]{rusin03}
Rusin, D., et al. 2003, \apj, 587, 143

\bibitem[Schneider et al.(2007)]{schneider07}
Schneider, D.~P., et al.\ 2007, \aj, 134, 102

\bibitem[Schneider et al.(2009)]{schneider09}
Schneider, D.~P., et al.\ 2009, in preparation

\bibitem[Schneider et al.(2006)]{schneider06}
Schneider, P., Kochanek, C.S., \& Wambsganss, J. 2006,
in Gravitational Lensing: Strong, Weak and Micro, Saas-Fee Advanced Course 33, G. Meylan,
P. North, P. Jetzer, eds., (Springer: Berlin)

\bibitem[Smith et al.(2002)]{smith02} 
Smith, A., et al. 2002, \aj, 123, 2121

\bibitem[Stoughton et al.(2002)]{stoughton02}
Stoughton, C., et al. 2002, \aj, 123, 485

\bibitem[Tucker et al.(2006)]{tucker06} 
Tucker, D. L., et al. 2006, Astron. Nachr, 327, 821


\bibitem[Wisotzki et al.(1993)]{wisotzki93}
Wisotzki, L., Koehler, T., Kayser, R., Reimers, D. 1993, A\&A, 278, L15

\bibitem[York et al.(2000)]{york00}
York, D. G., et al. 2000, \aj, 120, 1579

\end{thebibliography}
\end{document}